# Nano-Patterned Magnetic Edges in CrGeTe$_3$ for Quasi 1-D Spintronic Devices

*Avia Noah*[*1], Yishay Zur[1], Nofar Fridman[1], Sourabh Singh[1], Alon Gutfreund[1], Edwin Herrera[2], Atzmon Vakahi[3], Sergei Remennik[3], Martin Emile Huber[4], Snir Gazit[1,5], Hermann Suderow[2], Hadar Steinberg[1], Oded Millo[1], and Yonathan Anahory*[1]*

[1]The Racah Institute of Physics, The Hebrew University, Jerusalem, 9190401, Israel
[2]Laboratorio de Bajas Temperaturas, Unidad Asociada UAM/CSIC, Departamento de Física de la Materia Condensada, Instituto Nicolás Cabrera and Condensed Matter Physics Center (IFIMAC), Universidad Autónoma de Madrid, E-28049 Madrid, Spain
[3]Center for Nanoscience and Nanotechnology, Hebrew University of Jerusalem, Jerusalem, 91904, Israel
[4]Departments of Physics and Electrical Engineering, University of Colorado Denver, Denver, CO 80217, USA
[5]The Fritz Haber Research Center for Molecular Dynamics, The Hebrew University of Jerusalem, Jerusalem 91904, Israel

Email: avia.noah@mail.huji.ac.il, yonathan.anahory@mail.huji.ac.il

**Keywords:** Nano magnetism, Edge magnetism, Scanning SQUID microscopy, SQUID-on-tip, Magnetic imaging, van der Waals ferromagnet, CrGeTe$_3$

**Abstracts**

The synthesis of two-dimensional van der Waals magnets have paved the way for both technological applications and fundamental research on magnetism confined to ultra-small length scales. Edge magnetic moments in ferromagnets are expected to be less magnetized than in the sample interior because of the reduced amount of neighboring ferromagnetic spins at the sample edge. We recently demonstrated that CrGeTe$_3$ (CGT) flakes thinner than 10 nm are hard ferromagnets i.e. they exhibit an open hysteresis loop. In contrast, thicker flakes exhibit zero net remnant field in the interior, with hard ferromagnetism present only at the cleaved edges. This experimental observation suggests that a non-trivial interaction exists between the sample edge and interior. Here, we demonstrate that artificial edges fabricated by focus ion beam etching also display hard ferromagnetism. This enables us to write magnetic nanowire in CGT directly, and use this method to characterize the magnetic interaction between the interior and edge. The results indicate that the interior saturation and depolarization fields depend on the lateral dimensions of the sample. Most notably, the interior region between the edges of a sample narrower than 300 nm becomes a hard ferromagnet, suggesting an enhancement of the magnetic exchange induced by the proximity of the edges. Lastly, we find that the CGT regions amorphized by the gallium beam are non-magnetic, which introduces a novel method to tune the local magnetic properties of CGT films, potentially enabling integration into spintronic devices.

**Introduction:**

Low-dimensional magnetism[1–3] and specifically magnetically ordered van der Waals (vdW) materials[4–9], have attracted much interest in recent years. The timely and essential progress made now enables the study of unconventional magnetic phenomena with no direct counterpart in bulk 3D materials. Some examples include quantum spin chains[10–13], magnetic nanoparticles[14], and two-dimensional magnetic layers[15–20] These provide promising options for the experimental realization of phenomena, such as quantum criticality[21] and spin frustration[22], which have been the subject of numerous theoretical predictions. In particular, the intricate evolution of magnetic properties from bulk to thin exfoliated layers[9,23–25] offers insights into the physical origin of ferromagnetism (FM) in vdW materials, where anisotropy is thought to be the result of distinct inter-layer and intra-layer exchange interactions[4].

The vanishing remnant magnetization in zero field with increasing thickness is a phenomenon common to a number of vdW ferromagnetic materials.[25–27] For example, thin $CrGeTe_3$ (CGT) films ($d < 10$ nm), exhibit a net magnetization at zero applied field[4,27]. In contrast, using SQUID-on-tip (SOT) microscopy and in-situ magneto-transport measurements of $CGT/NbSe_2$ bilayers recent work demonstrated that the interior of thicker flakes ($d > 10$ nm) has zero remnant field, with hard FM appearing only at the sample edge[27]. This CGT edge magnetization is confined to a magnetic nanowire with a width and thickness of a few tens of nanometers. However, the physical mechanism causing edge magnetism remains to be identified. Modulation of the edge shape by nanofabrication could provide information about the role of the geometry in edge magnetization and about its magnetic interaction with the sample interior.

Beyond the interest in finding the underlying physical mechanism, edge magnetism could be applied in spintronic devices where magnetic nanowires serve for example as racetrack memory devices[28]. Here, we study edges nanofabricated by focused ion beam (FIB), and characterize them by scanning SOT microscopy[29,30]. Our key result is that magnetic edges can be directly written using a FIB. This capability allows us to examine magnetic edge confinement and the magnetic interaction between the sample interior and the edge. Our results indicate that when two edges are closer than 300 nm, the interior becomes a hard ferromagnet. In addition, we demonstrate that CGT regions amorphized by the gallium beam are non-magnetic, which introduces a novel method to tune the local magnetic properties in CGT films.

**Results:**

*Directly written magnetic edges*

In Fig. 1a, we present a schematic illustration of the experimental setup. CGT flakes with thicknesses ranging from 50 to 110 nm were exfoliated on top of a $SiO_2$ coated Si wafer. To create edges with controlled geometries, various shapes were etched out of the flakes using a $Ga^+$ FIB. Local magnetic field imaging $B_z(x, y)$ with a scanning SOT at 4.2 K was used to characterize the magnetic properties of the edges and surrounding areas. We estimate the spatial resolution of our images to approximatively 150 nm (see Methods and Supporting Note 1).

Fig. 1b depicts the topography of a CGT 50 nm thick flake from which a $2 \times 2$ μm$^2$ square-shaped hole was etched. The corresponding topographic line profile presented in Fig. 1c, demonstrates that the CGT was completely removed from this region. The topographic data was acquired *ex situ* in ambient conditions with a commercial atomic force microscope (AFM). Figures 1d,e shows $B_z(x, y)$ images of the same area as Fig. 1b. The field was ramped to $\mu_0 H_z = -200$ mT for a few seconds and subsequently returned to $\mu_0 H_z = 0$ before the SOT images were acquired (Fig. 1d). A similar field excursion was executed on a positive field ($\mu_0 H_z = 200$ mT) before the image shown in Fig. 1e was acquired. In both images, we measured a net magnetic signal at the edge of the square. The direction of the measured field at the edge is negative/positive after the respective field excursions at negative/positive applied magnetic field. In both cases, a disordered magnetic signal is observed $\sim 100$ nm away from the edge. To further analyze this results, Fig. 1f compares the magnetic signals cross-section $B_z(x)$ in locations indicated by dashed lines in Fig. 1d,e. The peaks observed in $B_z(x)$ at the edges (Fig. 1f blue and red curves) exhibit larger magnetic field values than the local field fluctuations in the disordered pattern observed far from the edges (Fig. 1f black curve). Furthermore, at the edge, the field direction is determined by the field history, whereas the interior average magnetization vanishes. We note that the width of the magnetic edge is limited by our tip size (175 nm) and the magnetic edge is certainly sharper than shown in the $B_z(x)$ profile. Our results are consistent with the magnetism found at cleaved edges of exfoliated CGT[27], thereby substantiating the ability to write magnetic edges in arbitrary shapes.

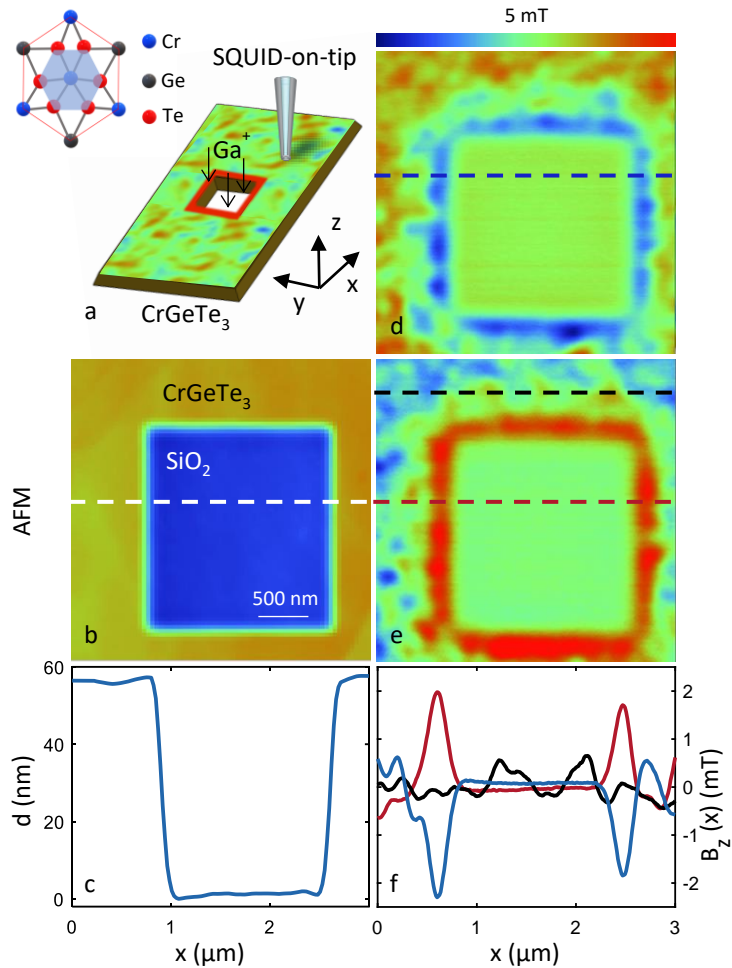

**Figure 1 SQUID-on-tip (SOT) images of nano-patterned edges in CrGeTe$_3$ at 4.2K. (a)** Schematic illustration of the measurement. (inset) Top view of the crystal structure. **(b)** Atomic force microscope (AFM) topographic image of the region of interest. **(c)** Topographic profile of the AFM measurement shown in **b**. **(d-e)** $B_z(x,y)$ images acquired at $\mu_0 H_z = 0$ after opposite field excursion $\mu_0 H_z = -200$ **d**, 200 **e** mT. **(f)** Line profile of the magnetic signal along the $x$-axis containing edges (blue and red dashed lines) and only the interior (black dashed line). All images are $3 \times 3$ μm$^2$ in size, pixel size 15 nm, acquisition time 5 min/image. The blue to red color scale represents lower and higher magnetic fields, respectively, with a shared scale of $B_z = 5$ mT.

*Amorphous CrGeTe₃*

To characterize all the possible effects of the FIB, we must also consider the potential amorphization of the CGT caused by the Ga$^+$ ion beam. We therefore configure the FIB to obtain a partially etched pattern that consists of three concentric annuli with different outer diameters (OD$_1$ = 2800 nm, OD$_2$ = 1200 nm, OD$_3$ = 350 nm) as depicted in the AFM image in Fig. 2a and illustrated schematically in Fig. 2b. The grooves visible in the AFM are not as deep as the thickness of the sample ($\sim 20$ nm $< d = 50$ nm). Fig. 2c shows the $B_z(x, y)$ image corresponding to the same area as the AFM image shown in Fig. 2a. To spatially resolve the resulting crystallographic structure, we prepared a scanning transmission electron microscopy (STEM) cross-section of the lamella corresponding to the region marked with the dashed line in Fig. 2c (see methods). The high-angle annular dark field (HAADF) image is shown in Fig. 2d. The crystalline material appears brighter in the STEM than the amorphized region (see also Supplementary Note 3 and Fig. S3). We note that the circles defining the annuli are not completely etched but that the remaining CGT is entirely amorphized. The area surrounding the etched area is also amorphized due to finite beam size effect. As a result, the CGT crystals are embedded in amorphous material. We also note that annulus #3 is almost completely amorphized while the other annuli retain a significant amount of crystalline material.

The $B_z(x, y)$ image (Fig. 2c) was acquired at $\mu_0 H_z = 0$ after a field excursion of $\mu_0 H_z = 200$ mT. We describe the magnetic features of this image starting from the frame of the image and going towards the center. In the region far from the annuli, the SOT image reveals the disordered magnetic domains averaging to zero magnetization. Next, we observe a ring color-coded in red, which corresponds to the edge of the sample interior in the vicinity of the largest amorphized ring. The amorphized ring appears in our SOT image as green and light blue. The annuli appear towards the center of the image. Fig. 2e presents the annuli stray field as a zoomed-in $B_z(x, y)$ image corresponding to the region marked in Fig. 2c and matching the region of the STEM cross-section in Fig. 2d. The outer red color-coded ring corresponds to annulus #1, which is fully magnetized at zero applied field. Further inwards is a smaller green color-coded ring, which is non-magnetic and corresponds to the etched region between annulus #1 and #2. Annulus #2 appears in the SOT image as a softer red color-coded ring. We note that annulus #3 is non-magnetic as the central area is color-coded in green at all measured applied fields.

By correlating the SOT $B_z(x, y)$ and STEM images, we conclude that all the amorphized regions, which include the three etched rings and annulus #3, are non-magnetic. Thus, we can define effective crystalline dimensions, for the width and thickness, $w_e$ and $d_e$. The effective crystalline dimensions of the outer annulus #1 are $w_e = 500$ nm and $d_e = 40$ nm, and those of the middle annulus #2 are $w_e = 100$ nm and $d_e = 10$ nm, which are comparable with the CGT domain size. This confinement gives rise to two distinct magnetic properties around the coercive field ($H_c$). At $H_z \sim H_c$, the annulus #1 is large enough to break into magnetic domains, while annulus #2 remains a single-domain and the magnetization of the entire area reverses abruptly (See Supporting Note 2).

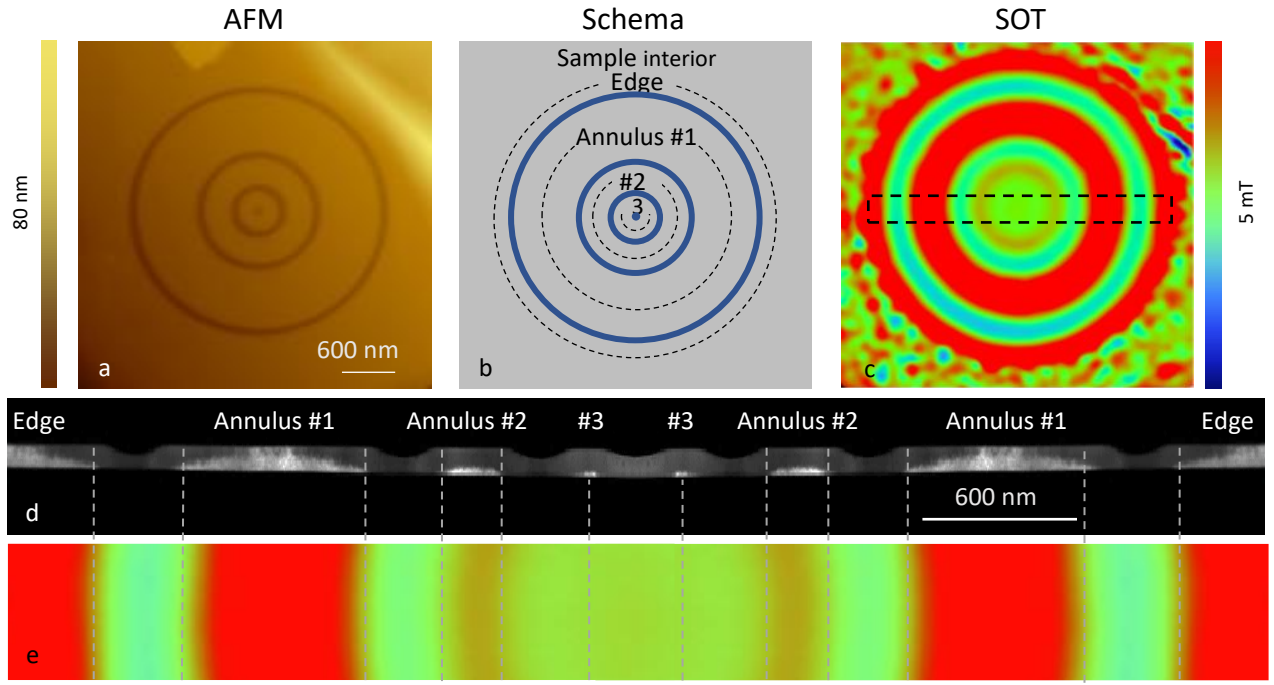

**Figure 2 SOT images of nano-patterned annuli in CrGeTe$_3$.** (**a**) Atomic force microscope (AFM) image of CrGeTe$_3$ (CGT) patterned using the focused ion beam (FIB) into annuli with outer diameters $OD_1 = 2800$ nm, $OD_2 = 1200$ nm, and $OD_3 = 350$ nm. (**b**) Schema of the annuli. The amorphized rings are shown in blue. The three Annuli and the outer edge are marked with dashed circles. (**c**) $B_z(x,y)$ images acquired at $\mu_0 H_z = 0$ after field excursion $\mu_0 H_z = 200$ mT. (**d**) Scanning transmission electron microscope (STEM) cross-sectional image measured along the black rectangle presented in **c**. The crystalline CGT appears in white, while the amorphized CGT appears in dark gray. The image was symmetrized around its center for clarity. (**e**) SOT image corresponding to the rectangle in **c** and matching the STEM cross-section in **d** at $\mu_0 H_z = 0$ mT. The gray dashed lines are a visual guide to facilitate the correlation between the crystalline CGT in the STEM **d** and SOT **e** image. The images are $4.5 \times 4.5$ μm² **c** and $3 \times 0.5$ μm² **e**, pixel size 18 nm, acquisition time 5 min/image. The blue to red color scale represents lower and higher magnetic fields, respectively, with a shared scale of $B_z = 5$ mT. The signal of annulus #1 intentionally saturates the color scale to allow the signal of annulus #2 to be visible on that scale.

*Edge-interior magnetic interaction*

In the next section, we use our ability to control nanoscale magnetic patterns in CGT to make a systematic study of the magnetic properties as a function of lateral dimensions. For this purpose, we use the Ga$^+$ FIB to fabricate 10 μm long CGT stripes with varying widths ($w_e$), as presented in the SEM image Fig. 3a. Fig. 3b represents a $B_z(x,y)$ image of the stripes acquired at $\mu_0 H_z = 0$ after a field excursion at $\mu_0 H_z = 200$ mT. The images of the wider stripes include two distinct magnetized edges (color-coded in red) separated by a zero-average magnetization in the stripe's interior (color-coded in green). However, below a certain critical width $w_c \sim 300$ nm, these two edges appear to merge to form a single magnetic domain. Fig. 3c represents a line-profile of the image shown in Fig. 3b along the $x$-axis ($B_z(x)$). We note that the $B_z(x)$ signal for the narrow stripe is four times larger than the signal at a single edge. This finding suggests that the stripe interior also becomes a hard ferromagnet because of the edge proximity.

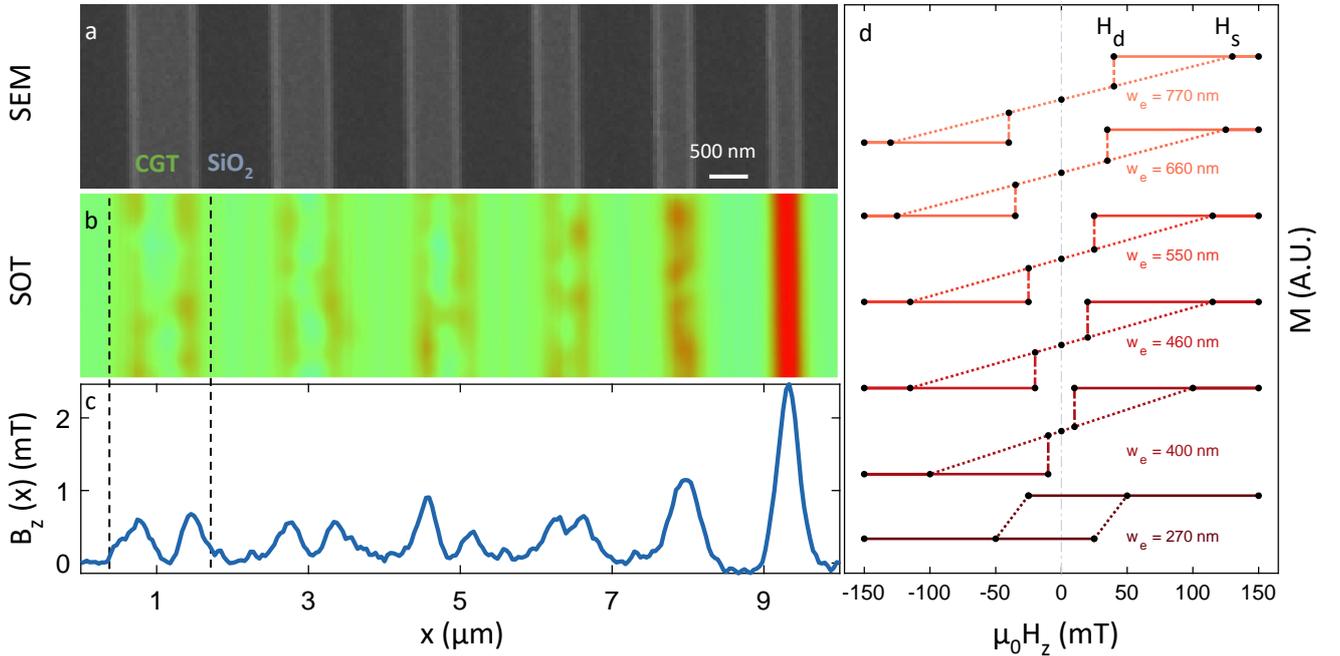

**Figure 3 - From 2-D to 1-D magnetic stripes.** (**a**) Scanning Electron Microscopy (SEM) image of CrGeTe$_3$ (CGT) with effective thickness $d_e = 50$ nm patterned into stripes with varying effective widths ($w_e$) and length of 10 μm. (**b**) SQUID-on-tip (SOT) magnetic image $B_z(x,y)$ acquired at $\mu_0 H_z = 0$ after positive field excursion to $\mu_0 H_z = 200$ mT. For stripes with $w_e > w_c$, two distinct magnetized edges (red color scale) separated by a zero average magnetization in the stripe's interior (color-coded in green). For stripes with width $w_e < w_c$ (right stripe), the two edges appear to merge and form a single magnetic domain. (**c**) Line profile of the magnetic signal along the x-axis of the image in **b**. The line profile was averaged over 45 pixels. (**d**) Sketched magnetization curves drawn from $B_z(x,y)$ acquired on stripes with different widths. Dashed lines are a guide to the eye connecting the saturated fields $H_s$ and the demagnetization field $H_d$. The fields at which the images were taken are marked with black dots. The SOT image is $2.5 \times 10$ μm², pixel size 40 nm, acquisition time 5 min/image. The blue to red color scale represents lower and higher magnetic field, respectively, with a scale of $B_z = 5$ mT.

To understand this observation quantitatively, we carried out magnetostatic simulations assuming a magnetization of 3 μ$_B$/Cr and a unit cell volume of 0.83 nm³.[31] The tip-to-sample distance (170 nm) can be obtained by coupling the tip to a tuning fork to sense the surface[32]. The geometry of the sample is obtained from the HAADF images considering that the amorphous CGT is non-magnetic (Fig. 4a,b). The stripe cross-section is trapezoid-shaped and marked with green dashed lines. Given this geometry, we define the effective width as $w_e = \frac{w_{base} + w_{top}}{2}$. The STEM resolves an effective dimension of $w_e = 460$ nm (Fig. 4a) and 270 nm (Fig. 4b) with an effective thickness $d_e = 50$ nm.

For a wider stripe ($w_e = 460$ nm, Fig. 4a), we model the edge by assuming a right-angled triangle cross-section with area $L \times d/2 = 2000$ nm² (Fig. 4a). The simulated field distribution emanating from such a triangular cross-section edge (Fig. 4f) is in good agreement with the measured SOT image (Fig. 4g). Thus, the simulation confirms a magnetic edge width of a few tens of nanometers. The minor discrepancy observed between the SOT image and simulation may be due to local variations in the edge roughness or may be a consequence of the influence of the magnetic domains present in the bulk. Fig. 4d presents the same calculation executed for the narrow stripe ($w_e = 270$ nm), by modeling the edges as a triangle cross-section of $L \times d/2 = 2500$ nm². In contrast to the wider stripe, this yields a poor agreement between the simulation (Fig. 4h) and the SOT results (Fig. 4i), where the simulated signal magnitude is smaller than the experimental data by a factor of four. To obtain good agreement, we need to assume that the entire stripe is magnetized as illustrated in Fig. 4e and simulated in Fig. 4j. We therefore conclude that the edges enhance the exchange interaction in the sample interior, resulting in a proximity-induced hard ferromagnetic state. The typical decay length of such interaction can be estimated as about $w_c = 300$ nm for $d_e = 50$ nm and this defines the critical width $w_c$ for the emergence of hard FM in the sample interior.

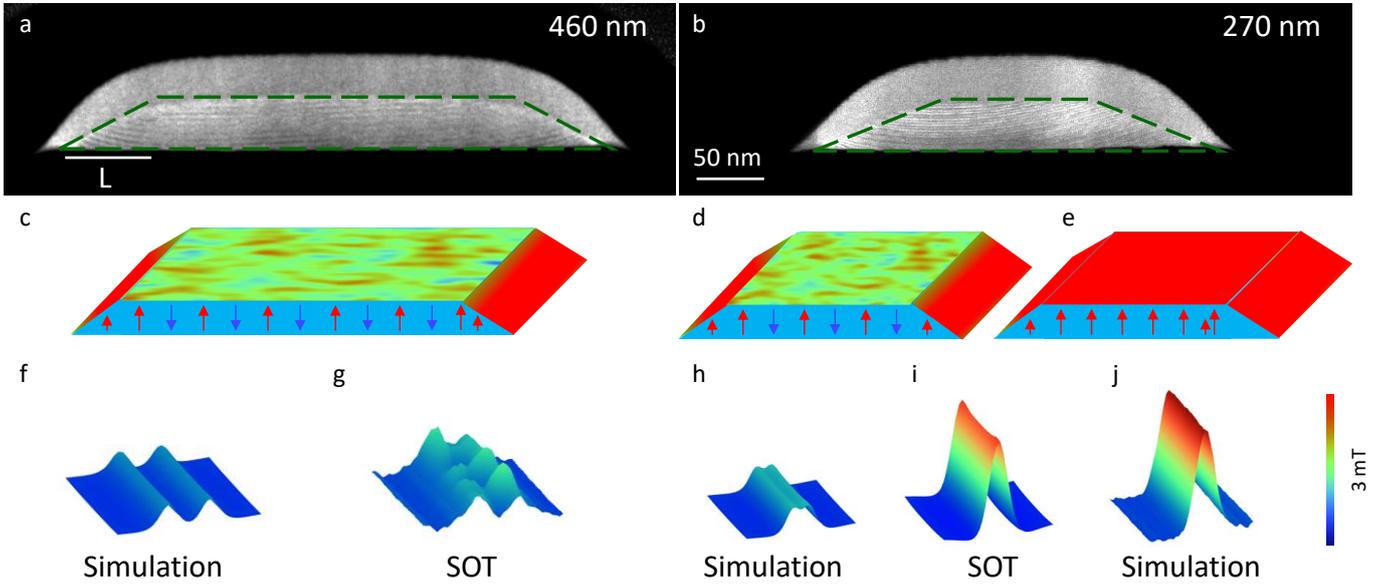

**Figure 4 STEM images of the CrGeTe₃ stripes and magnetostatic simulations.** (**a-b**) Scanning transmission electron microscope (STEM) cross-sectional images measured in the middle of the stripe with effective widths $w_e = 460$ nm **a,** and $w_e = 270$ nm **b**. The effective crystalline CrGeTe₃ is marked with green dashed lines. (**c-e**) Schematic illustration of the local magnetic structure at the edges and interior for different stripe width. In panels **c-d,** the edges but not the interior of the sample retain their magnetization. In panel **e,** the whole stripe is a hard ferromagnet. (**f-j**) A comparison between the SQUID-on-tip (SOT) images and magnetostatic simulations. (**f,h**) Simulations of the stripe magnetization resulting from the magnetized edges for a triangular cross-section, marked by red in panel **c,d**. (**j**) Simulations of the stripe magnetization assuming that the whole stripe is magnetized, marked by red in panel **e**. (**g,i**) $B_z(x,y)$ SOT images of the stripes in **a,b**. The blue to red color scale represents lower and higher magnetic fields, respectively, with a shared scale of 3 mT.

The evolution of the magnetic profile seen in Fig. 3b,c suggests that there is an abrupt transition from a soft to a hard ferromagnet in the sample interior. However, more precise magnetic characterization reveals that the transition is gradual. By measuring $B_z(x,y)$ as a function of the applied field $H_z$, we can extract the width dependence of the saturation ($H_s$) and demagnetization ($H_d$) fields on the sample interior. Fig. 3d, summarizes the values of $H_s$ and $H_d$ found for each stripe, with the magnetization hysteresis curves marked as a dashed line connecting the dots corresponding to $H_s$ and $H_d$ as a visual guide. For $w > w_c$, $H_d$ and $H_s$ grow with $w$, which produces a bowtie hysteresis curve with no remnant field in the sample interior. For $w < w_c$, the demagnetization field crosses zero and becomes the coercive field $H_c$, which results in an open hysteresis loop. In this case, the stripe breaks into magnetic domains at the coercive field $\mu_0 H_c = \pm 25$ mT and the magnetization saturates at $\mu_0 H_S = \pm 50$ mT. This set of measurements indicates that the continuous effect of magnetic confinement before the transition causes the sample interior to become a hard ferromagnet. It is important to note that for $w < w_c$, the sample interior is still sufficiently large to accommodate magnetic domains at the coercive field. Thus, the observed transition does not coincide with a single-domain transition although $w_c = 300$ nm is comparable with the magnetic domain size ($\sim 100$ nm).

**Discussion**

One plausible underlying physical mechanism for edge magnetism is related to the wedge shape seen in cleaved edges[27]. Given that thin flakes ($d < 10$ nm) are hard ferromagnets and considering that part of the wedge must be thinner than 10 nm, this mechanism appears relevant for cleaved edges. Since etched edges possess a similar angle that obtained upon cleaving (20° to 30°), the same mechanism could also explain edge magnetism in etched edges.

Another potential explanation for edge magnetism in etched edges could be the $Ga^+$ contamination. Cross-sectional and energy-dispersive X-ray spectroscopy (EDS) measurements of the annuli (Fig. 2) and the stripes (Fig. 3) addressing the Ga contamination and CGT oxidation are presented in Supporting figures 4-6. Importantly, in the crystalline regions, the Ga concentration is uniform and below the background level $\sim 1.5$ at %. The highest Ga concentration (15 at %) is uniformly distributed on the surface of the amorphized region where the concentration is 2.5 at %. We note that regions with the highest Ga concentration are amorphized and were found to be non-magnetic. This observation suggests that if the presence of Ga has an effect, it would be to hinder magnetism rather than enhancing it. We note that the uniformity of the Ga distribution suggests that the edge magnetism cannot be explained by the Ga concentration profile. Oxygen contamination is spread uniformly near the CGT surface. No measurable amount of oxygen was observed 5 nm below the surface.

Strain should also be considered as one of the mechanisms to induce edge magnetization. It was shown that strain can enhance magnetism in CGT[33,34]. It is plausible that some strain appears at low temperature at the interface between the amorphous and crystalline CGT regions. The last mechanism that we can consider is related to the in-plane dangling bond. This mechanism was previously excluded[27] since no magnetic edge was found at the step edge between two terraces of a single-flake where in-plane dangling bonds would be expected. In addition, for nanofabricated edges, the magnetic edge is embedded in amorphous CGT, which should reduce the number of in-plane dangling bonds. Since the edge magnetism observed for these embedded edges is similar to that of an exfoliated sample, we consider this mechanism to be unlikely.

**Conclusion**

In conclusion, we have demonstrated that quasi-1D magnetic edges can be directly written to form arbitrary shapes by using the FIB. That capability allows us to measure the effect of lateral confinement. In particular, we have shown that when two edges are separated by less than 300 nm, the whole sample becomes a hard ferromagnet. This suggests that geometry can influence the microscopic exchange interaction, strengthening ferromagnetic exchange over large distances. In addition, we report that an amorphous CGT material is non-magnetic, which introduces an additional method to control the local magnetism. The directly written magnetic structure could be useful in devices that require very narrow magnetic channels, and we believe that the new method will have great potential for applications and fundamental research in confined magnetism and serve as a building block for spintronic devices.

## Methods

**Sample fabrication:**

CrGeTe$_3$ (CGT) crystals were grown using the flux method[35]. CGT samples were fabricated using the dry transfer technique, which was carried out in a glovebox with an argon atmosphere. The CGT flakes were cleaved using the scotch tape method, and exfoliated on commercially available Gelfilm from Gelpack[27]. For the SQUID-On-Tip (SOT) measurements, a CGT flake was transferred on a SiO$_2$ substrate. The various shapes were etched out of the CGT flakes with a Ga$^+$ focused ion beam (FIB). The flakes were ~50-110 nm thick as determined by atomic force microscopy and scanning transmission electron microscopy (STEM) measurements.

To fully etch the 110 nm thick CGT flake shown in Fig. 3, we utilized a Ga$^+$ ion beam operating at 30 kV and a 790 pA current. We etched a rectangular area of 10 µ$m^2$ during 10 seconds, which results in a fluence of ~$10^{17}$ Ga$^+$/cm$^2$. For the under-etched regions, such as the annuli rings shown in Fig. 2, we employed a Ga$^+$ ion beam at 30 kV and a 1.1 pA current. The largest ring had a diameter of 3 µm, and its width was estimated to the Ga beam profile (200 nm). In this case, we etched an area of $0.2 \times 2\pi \times 1.5 =$ ~1.9 µ$m^2$ for 10 seconds, which resulted in a fluence of ~$10^{15}$ Ga$^+$/cm$^2$. To fully etch the 50 nm thick CGT flake shown in Fig. 1, we utilized a Ga$^+$ ion beam at 30 kV and a 7.7 pA current. We etched a square area of 4 µ$m^2$ for 18 seconds, which resulted in a fluence of ~$10^{16}$ Ga$^+$/cm$^2$.

**Scanning SQUID-On-Tip microscopy:**

The SOT was fabricated using self-aligned three-step thermal deposition of Pb at cryogenic temperatures, as described previously[29]. The measurements were performed using tips with effective SQUID loop diameters ranging from 145 to 175 nm. Figure S1 shows the measured quantum interference pattern of one of the SOTs used for this work, which has an effective diameter of 145 nm and a maximum critical current of 110 µA. The asymmetric structure of the SOT gives rise to a slight shift of the interference pattern, resulting in good sensitivity in zero field. All measurements were performed at 4.2 K in a low pressure He of 1 to 10 mbar. All images were acquired with a constant distance between the tip and sample (170 nm). In these conditions, the magnetic signal measured, which is on the order of 0.5-3 mT is much larger than any possible parasitic influence related to the varying topography. That parasitic signal is estimated to be much smaller than our magnetic signal (<0.01 mT).

**Sample characterization:**

High-resolution scanning electron microscope (SEM) cross-section lamellas were prepared and imaged by Helios Nanolab 460F1 Lite focused ion beam (FIB) - Thermo Fisher Scientific. The site-specific thin lamella was extracted from the CGT patterns using FIB lift-out techniques[36]. STEM and Energy-Dispersive X-ray Spectroscopy (EDS) analyses were conducted using an Aberration Prob-Corrected S/TEM Themis Z G3 (Thermo Fisher Scientific) operated at 300 KV and equipped with a high-angle annular dark field detector (Fischione Instruments) and a Super-X EDS detection system (Thermo Fisher Scientific).

**Supporting Information:**
The supporting information contains the following: additional experimental details and discussion such as information on the SOT parameters, SOT images at the coercive field of patterned annuli, characterization of the amorphous and crystalline regions, and EDS measurements of the CGT stripes and annuli.

**Acknowledgements:**

We would like to thank A. Capua, Y. Paltiel and B. Yan for fruitful discussions. Devices for this project were fabricated at the Hebrew University center for Nanoscience. This work was supported by the European Research Council (ERC) Foundation grant No. 802952. The international collaboration on this work was fostered by the EU-COST Action CA21144. H. Steinberg acknowledges funding provided by the DFG Priority program grant 443404566 and Israel Science Foundation (ISF) grant 861/19. O. Millo is grateful for support from the Academia Sinica – Hebrew University Research Program, the ISF grant no. 576/21, and the Harry de Jur Chair in Applied Science. S. Gazit acknowledges support from the ISF grant No. 586/22. H. Suderow and E. Herrera acknowledge support from the Spanish State Research Agency (PID2020-114071RB-I00, CEX2018-000805-M) and the Comunidad de Madrid through the NANOMAGCOST-CM program (Program No.S2018/NMT-4321).

**Author Contributions:**

Y.A., A.N., O.M., S.G. conceived the experiment.
E.H. and H.S. synthesized the CGT crystals.
A.N. and N.F. carried out the scanning SOT measurements.
Y.Z., and A.N. computed the simulation.
Y.A., S.S., A.V., H. Steinberg, and A.N. fabricated and characterized the CGT devices.
A.V. and S.R. carried out the TEM measurements.
A.N. analyzed the data.
Y.A., A.N., and A.G. constructed the scanning SOT microscope.
M.E.H. developed the SOT readout system.
A.N., O.M., and Y.A. wrote the paper with contributions from all authors.
Notes: The authors declare no competing financial interest.

**List of abbreviations:**

Atomic force microscopy (AFM)
Coercive Field ($H_c$)
Critical width ($w_c$)
CrGeTe$_3$ (CGT)
Demagnetization field ($H_d$)
Energy-Dispersive X-ray Spectroscopy (EDS)
Effective thickness ($d_e$)
Effective width ($w_e$)
Ferromagnetism (FM)
Focused Ion Beam (FIB)
High-angle annular dark field (HAADF)
OD (outer diameter)
Saturation field ($H_s$)
Scanning Transmission Electron Microscopy (STEM)
SQUID-on-tip (SOT)
Thickness ($d$)
van der Waals (vdW)

# Supporting Information

# Nano-Patterned Magnetic Edges in CrGeTe$_3$ for Quasi 1-D Spintronic Devices

*Avia Noah*[1], Yishay Zur[1], Nofar Fridman[1], Sourabh Singh[1], Alon Gutfreund[1], Edwin Herrera[2], Atzmon Vakahi[3], Sergei Remennik[3], Martin Emile Huber[4], Snir Gazit[1,5], Hermann Suderow[2], Hadar Steinberg[1], Oded Millo[1], and Yonathan Anahory*[1]*

[1]The Racah Institute of Physics, The Hebrew University, Jerusalem, 91904, Israel
[2]Laboratorio de Bajas Temperaturas, Unidad Asociada UAM/CSIC, Departamento de Física de la Materia Condensada, Instituto Nicolás Cabrera and Condensed Matter Physics Center (IFIMAC), Universidad Autónoma de Madrid, E-28049 Madrid, Spain
[3]Center for Nanoscience and Nanotechnology, Hebrew University of Jerusalem, Jerusalem, 91904, Israel
[4]Departments of Physics and Electrical Engineering, University of Colorado Denver, Denver, CO 80217, USA
[5]The Fritz Haber Research Center for Molecular Dynamics, The Hebrew University of Jerusalem, Jerusalem 91904, Israel

Email: avia.noah@mail.huji.ac.il, yonathan.anahory@mail.huji.ac.il

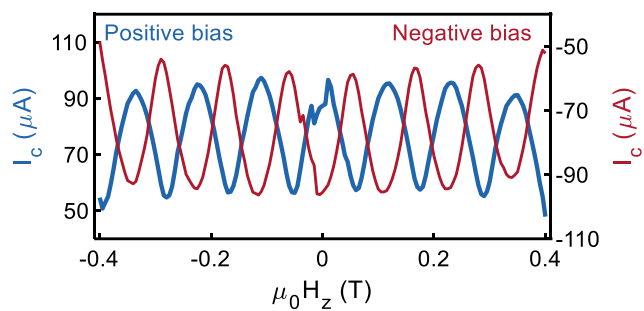

**Figure S1. Quantum interference pattern of the SQUID-on-tip (SOT).** The critical current $I_c$ of one of the SOT's used in this work as a function of the applied out-of-Plane field $H_z$. Blue: Positive bias, red: Negative bias. The period of 120 mT of the quantum interference corresponds to an effective diameter of 145 nm of the SOT.

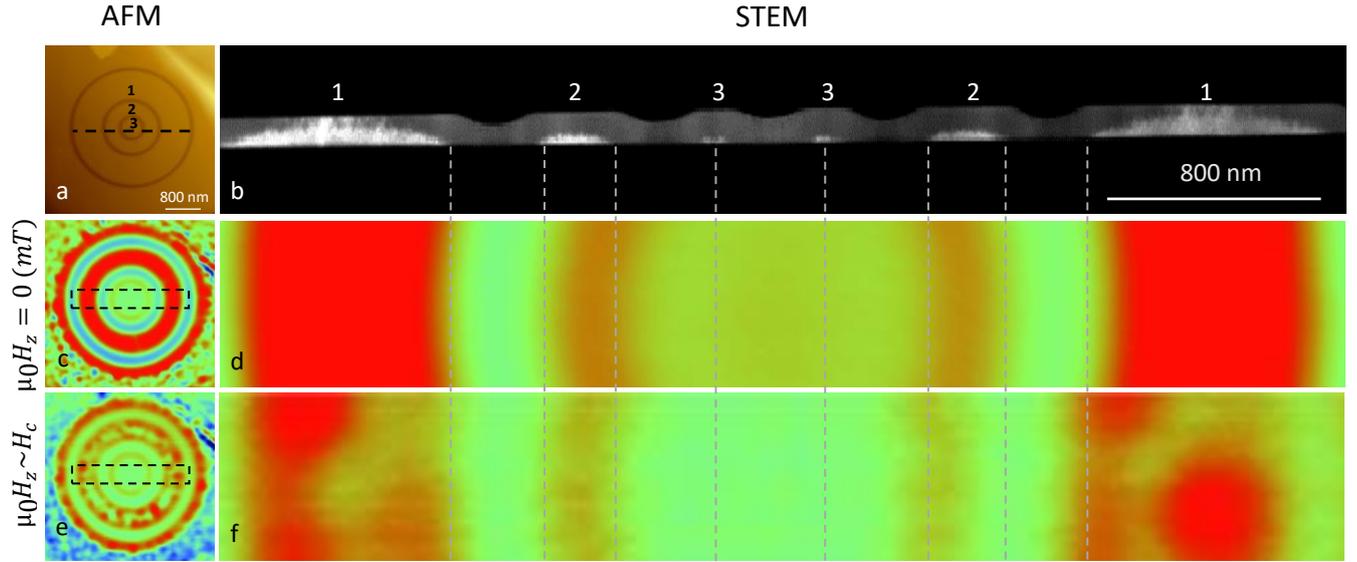

**Figure S2. SOT images of patterned annuli in CrGeTe₃.** (a) AFM image of CrGeTe₃ patterned using FIB into annuli with outer diameters $OD_1 = 2800$ nm, $OD_2 = 1200$ nm and $OD_3 = 350$ nm. (b) HAADF STEM cross-sectional image measured along the black line presented in **a**. The crystalline CGT appears in white, while the amorphized CGT appears in dark gray. (c) $B_z(x,y)$ images acquired at $\mu_0 H_z = 0$ after field excursion $\mu_0 H_z = 200$ mT. (d) SOT image corresponding to the rectangle in **c** and matching the STEM cross-section in **b**. (e) $B_z(x,y)$ images acquired at $0 < \mu_0 H_z < \mu_0 H_c$. (f) SOT image corresponding to the rectangle in **e**. At $H_z \sim H_c$, the annulus #1 is break into magnetic domains, while annulus #2 stays as a single domain. The dashed lines are a guide to the eye of the crystalline annuli. The images are $4.5 \times 4.5$ μm² **c,e** and $3 \times 0.5$ μm² **d,f**, pixel size 18 nm, acquisition time 5 min/image. The blue to red color scale represents lower and higher magnetic field, respectively, with a shared scale of $B_z = 5$ mT.

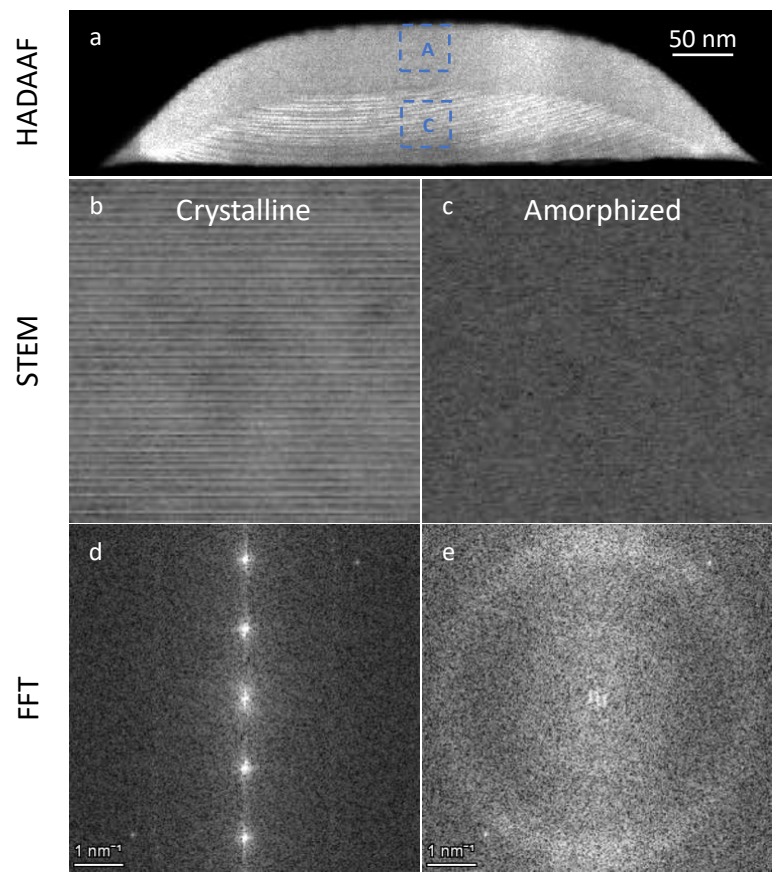

**Figure S3. Zoomed-in STEM image of the 270 nm stripe in figure 4. (a)** High-angle annular dark field (HAADF) image of the stripe cross section. **(b-c)** Zoomed-in of the crystalline region – bottom rectangle in **a** (b) and the amorphized region – top rectangle in **a** (c). **(d-e)** FFT of the images in b-c respectively, showing a Fourier transform of a crystal structure along the c axis (d) and of an amorphized structure (e).

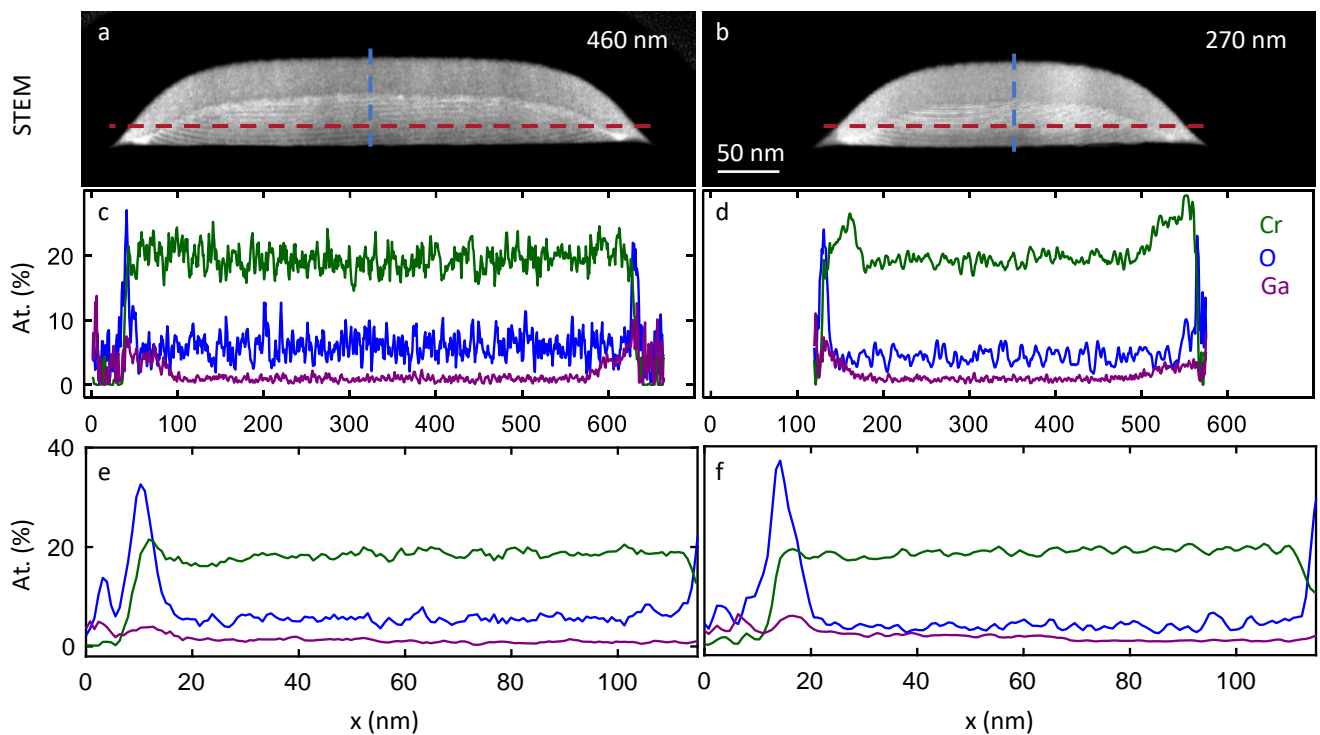

**Figure S4. Cr, O, and Ga EDS line-scan of stripes in figure 4. (a-b)** High-angle annular dark field (HAADF) image of the stripe cross section. **(c-f)** Energy-Dispersive X-ray Spectroscopy (EDS) line scan, showing the relative amount of Cr, O, and Ga in a horizontal (c,d) and vertical (e,f) cross sections of the stripes, in atomic percent.

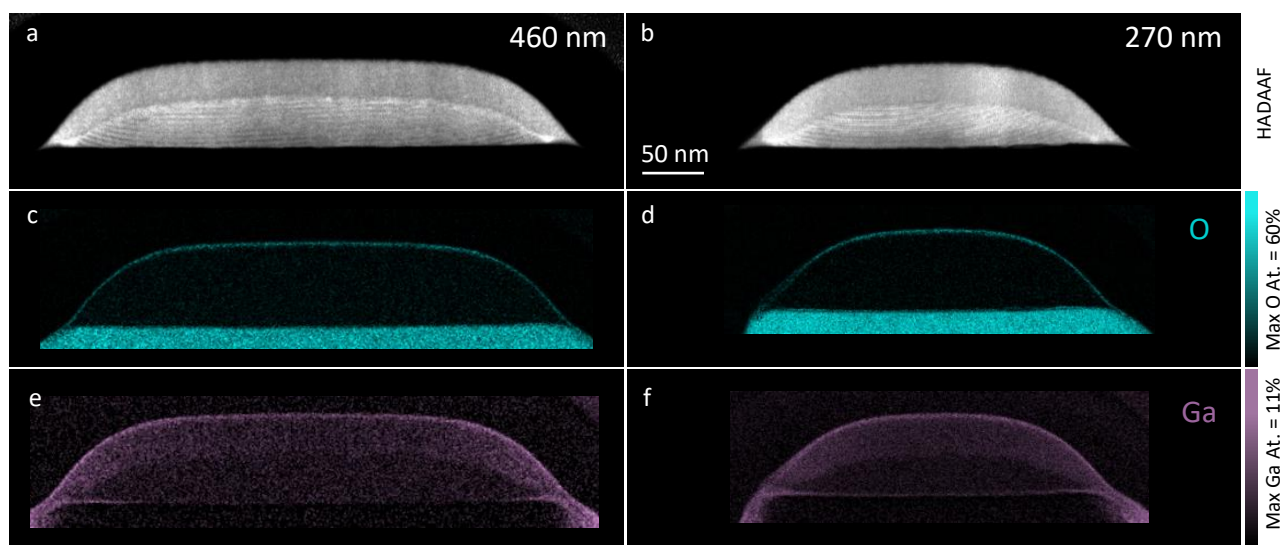

**Figure S5. O, and Ga EDS maps of stripes in figure 4. (a-b)** High-angle annular dark field (HAADF) image of the stripe cross section. **(c-f)** Energy-Dispersive X-ray Spectroscopy (EDS) map, showing the relative amount of O (c,d), and Ga (e,f) in a cross sections of the stripes, in atomic percent.

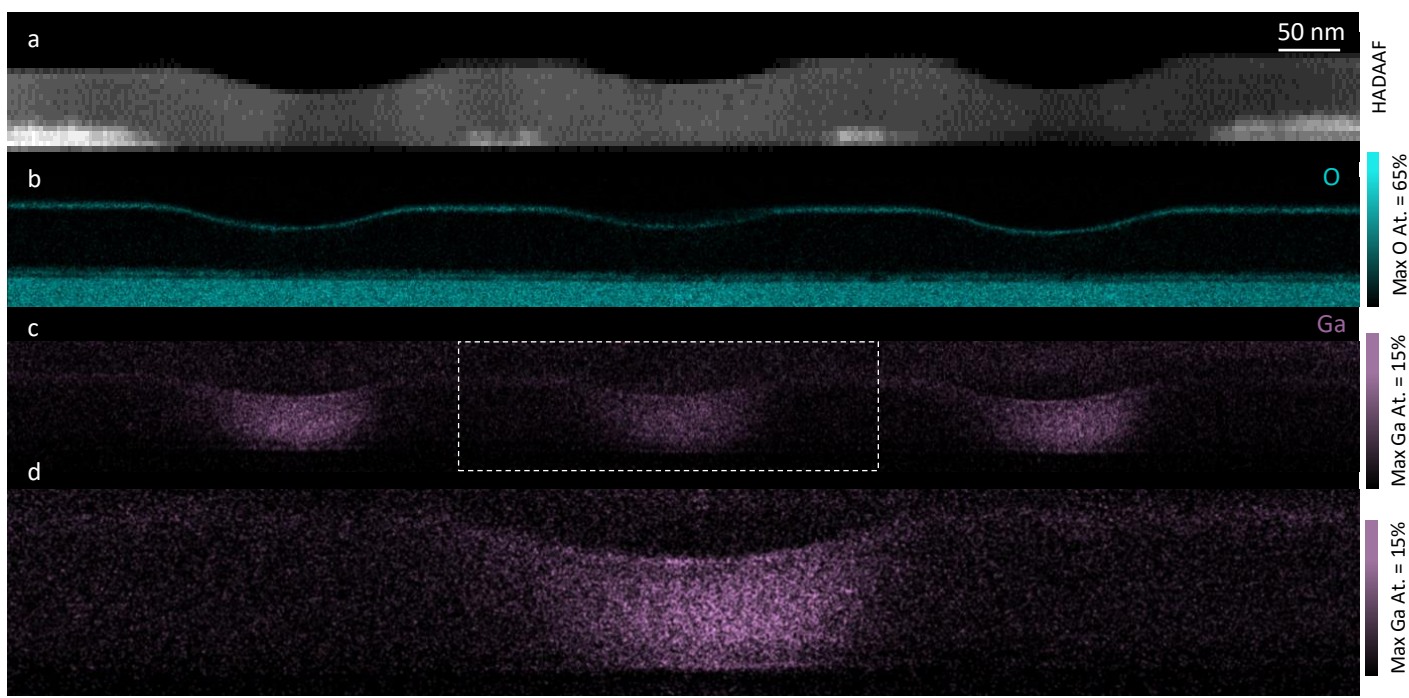

**Figure S6. O, and Ga EDS maps of annuli in figure 2. (a)** High-angle annular dark field (HAADF) image of the center annuli cross section. **(b-d)** Energy-Dispersive X-ray Spectroscopy (EDS) line scan, showing the relative amount of O (b) and Ga (c-d) in a cross section of the annuli, in atomic percent. (**d**) Zoomed-in image of the dashed rectangle in c.

**Supporting Note 1: SOT fabrication and characterization**

Figure S1 shows the quantum interference pattern typical for a SQUID. The SOT response to magnetic field is defined as the derivative of that pattern. To get good images, the SOT has to be field biased in a region where the interference pattern is linear, and one must avoid the regions where the response is zero (blind spots). The field period of this pattern is determined by the SQUID loop diameter so that each period represents a magnetic flux equal to the quantum of flux $\phi_0 = h/2e \approx 20.67$ G$\mu$m$^2$. This means that a smaller SQUID loop will exhibit a large field period with larger linear regions but with larger blind spots. Depending on the experimental requirements, optimal SQUID loop size is chosen. To mitigate the effect of blind spots, we use the fact that our SOTs are often composed of asymmetric junctions. This makes the interference pattern to have a different field offset for a different direction of the current running through the SQUID (Figure S1, blue and red curves).

The magnetic length scale in CrGeTe$_3$ is on the order 100 nm or smaller. That implies a SQUID loop with a diameter below 100 nm. However, having a SQUID with a loop too small, of 50 nm for example, would yield a blind spot of +/- 0.4 T around zero field. That would be problematic to get good images below the saturation field (0.13 T). For this reason, all the SOTs used in this work have a SQUID loop of about 145 to 175 nm.

**Supporting Note 2: SOT images of patterned annuli in CrGeTe₃**

Here we compare between the annuli magnetizations. As mentioned in the manuscript, the CGT sample were partially etched using FIB and consists of three concentric annuli with different outer diameters (OD$_1$ = 2800 nm, OD$_2$ = 1200 nm, OD$_3$ = 350 nm) as depicted in the AFM image (Figure S2a). The STEM cross-section lamella in Figure S2b corresponding to the region marked with the dashed line in Figure S2a. Figure S2c shows the $B_z(x, y)$ image corresponding to the same area as the AFM image. The image was acquired at $\mu_0 H_z = 0$ after field excursion of $\mu_0 H_z = 200$ mT. To describe the annuli magnetization, we present in Figure S2d a zoomed-in $B_z(x, y)$ image corresponding to the region marked in Figure S2c. The red color-coded ring corresponds to annulus #1, which is fully magnetized at zero applied field. The next to that feature is a smaller nearly green color-coded ring, which is non-magnetic. This corresponds to the region between annulus #1 and #2. Annulus #2 is then visible as softer red color-coded. We note that annulus #3 is non-magnetic as the central area is green color-coded at all measured applied fields.

Now, we discuss the difference between the magnetic properties of annulus #1 and #2. The effective crystalline dimensions of annulus #1 are $w_e = 500$ nm and $d_e = 40$ nm. For annulus #2 the dimensions are $w_e = 100$ nm and $d_e = 10$ nm. This confinement gives rise to two distinct magnetic properties around the coercive field. In Figure S2e, we present the $B_z(x, y)$ image corresponding to the same area as Figure S2c but at $H_z \sim H_c$. In Figure S2f we present a zoomed-in $B_z(x, y)$ image of the marked area (same area as Figure S2d). Annulus #1 and annulus #2 hold the magnetization at zero field. However, at $H_z \sim H_c$, the annulus #1 is sufficiently large to break into magnetic domains, while annulus #2 stays as a single domain, and the magnetization of the entire area reverses abruptly. We attribute the single domain effect to the lack of space to accommodate magnetic domains.

**Supporting Note 3: Characterization of the amorphous and crystalline regions.**

To distinguish between the crystalline and amorphized regions we present in figure S3 high-magnification images of the 270 nm STEM image with the corresponding FFT analysis. In the crystalline part, the atomic layers are clearly resolved as seen in pristine samples and the FFT peaks corresponding to the interatomic layer are visible. In the amorphous region, we see a circular pattern in the FFT in agreement with an amorphous phase.